\documentclass[12pt]{article}

\usepackage{axodraw}
\usepackage{epsf}
\usepackage[dvips]{graphicx}
\begin{document}

\begin{center}
{\bfseries PROMPT PHOTON PRODUCTION WITH $K_T-$FACTORIZATION }

\vskip 5mm

A.V. Lipatov$^{1 \dag}$, N.P. Zotov$^{2 \dag}$ 

\vskip 5mm

{\small
 {\it
D.V. Skobeltsyn Institute of Nuclear Physics,
M.V. Lomonosov Moscow State University 
}
\\
$1\dag$ {\it
E-mail: lipatov@theory,sinp.msu.ru,
$2\dag$ zotov@theory.sinp.msu.ru
}}
\end{center}

\vskip 5mm

\begin{center}
\begin{minipage}{150mm}
\centerline{\bf Abstract}
 We consider the prompt photon production at modern high energy 
 colliders in the framework of $k_T-$factorization  
approach. We compare our theoretical predictions with recent 
experimental data at HERA and Tevatron, empahasizing the distinction
 between our theoretical predictions and the results of NLO QCD 
calculations. Finally, we extrapolate our predictions 
to LHC energies.
\end{minipage}
\end{center}

\vskip 10mm

\section{Introduction}
 It is well known that production of prompt (or direct) photons at
high energies has provided a direct probe of the hard subprocess 
dynamics,
since produced photons are largely insensitive to the final-state
hadronization effects.  Usually photons are called "prompt" if 
they are coupled to the interacting quarks. In the framework of 
QCD these photons in $ep$ collisions can be produced via direct
$\gamma q \to \gamma q$ and resolved $ g q \to \gamma q$ 
production
mechanisms. The last-named mechanism is dominant production one 
for the prompt photons in $pp$ collisions. It is clear that cross 
section of such processes is sensitive to the parton distributions 
in a proton and a photon. Also observed final state photons may 
arise from so called fragmentation
processes, where a quark or gluon are transformed into 
$\gamma$. However, the isolation criterion which is usually 
introduced in experimental analyses
substantially reduces the fragmentation component (see, for 
example, Ref.~[1]).
The prompt photon production in $ep$ and $pp$ collisions has been
studied in a number of experiments at HERA~[2 - 4] and 
Tevatron~[5, 6].

In $pp$ collisions it was found~[5, 6] that the 
shape of the measured cross sections as a
function of photon transverse
energy $E_T^\gamma$ is poorly described by next-to-leading order 
(NLO)
QCD calculations: the observed $E_T^\gamma$ distribution is
steeper than the predictions of perturbative QCD.
These shape differences lead to a significant disagreement
in the ratio of cross sections calculated at different 
center-of-mass
energies $\sqrt s = 630$ GeV and $\sqrt s = 1800$ GeV as a 
function of scaling variable
$x_T = 2 E_T^\gamma/\sqrt s$. The disagreement in the
$x_T$ ratio is difficult to explain with conventional
theoretical uncertainties connected with scale dependence and
parametrizations of the parton distributions~[5, 6]. The origin of
the disagreement has been ascribed to the effect of initial-state
soft-gluon radiation [7, 8].
It was shown that observed discrepancy can be reduced by
introducing some additional intrinsic transverse momentum $k_T$ of 
the incoming partons, which is usually assumed to have a
Gaussian-like distribution~[8, 9]. However,
the average value of this $k_T$ increases from
$\langle k_T \rangle \sim 1$ GeV to more than
$\langle k_T \rangle \sim 3$ GeV in  hard-scattering processes as
the $\sqrt s$ increases from UA6 to Tevatron energies~[8, 10].

The treatment of $k_T-$enhancement
in the inclusive prompt photon hadroproduction at Tevatron
proposed in  Ref.~[11], based on
the $k_T$-factorization QCD approach, suggests possible 
modifications of the above simple $k_T$ smearing picture.
The unintegrated parton distributions in a proton were obtained
using the KMR formalism and the role of the both non-perturbative
and perturbative components of partonic transverse momentum $k_T$ 
in describing of
the observed $E_T^\gamma$ spectrum was investigated.
However, the KMR unintegrated parton densities were
obtained in the double leading logarithmic approximation (DLLA) 
only. Also in these calculations the usual on-shell matrix
elements of hard partonic subprocesses were
evaluated with precise off-shell kinematics.

 In our papers [12, 13] we have applyed the KMR method to 
obtain  the unintegrated quark and gluon distributions  
$f_a(x,{\mathbf k}_T^2,\mu^2)$ in a proton and a photon
independently from other authors. Then we have studied
prompt photon production at HERA and Tevatron in more detail.
Here we present some of these results, which demonstrate our 
specific predictions in comparison with the results of
NLO QCD calculations.
    
\section{Prompt photon production at HERA}
\begin{figure}
\hspace*{0mm}
\begin{minipage}[h]{.49\textwidth}
\includegraphics[height=9.0cm,width=8.0cm,angle=0]{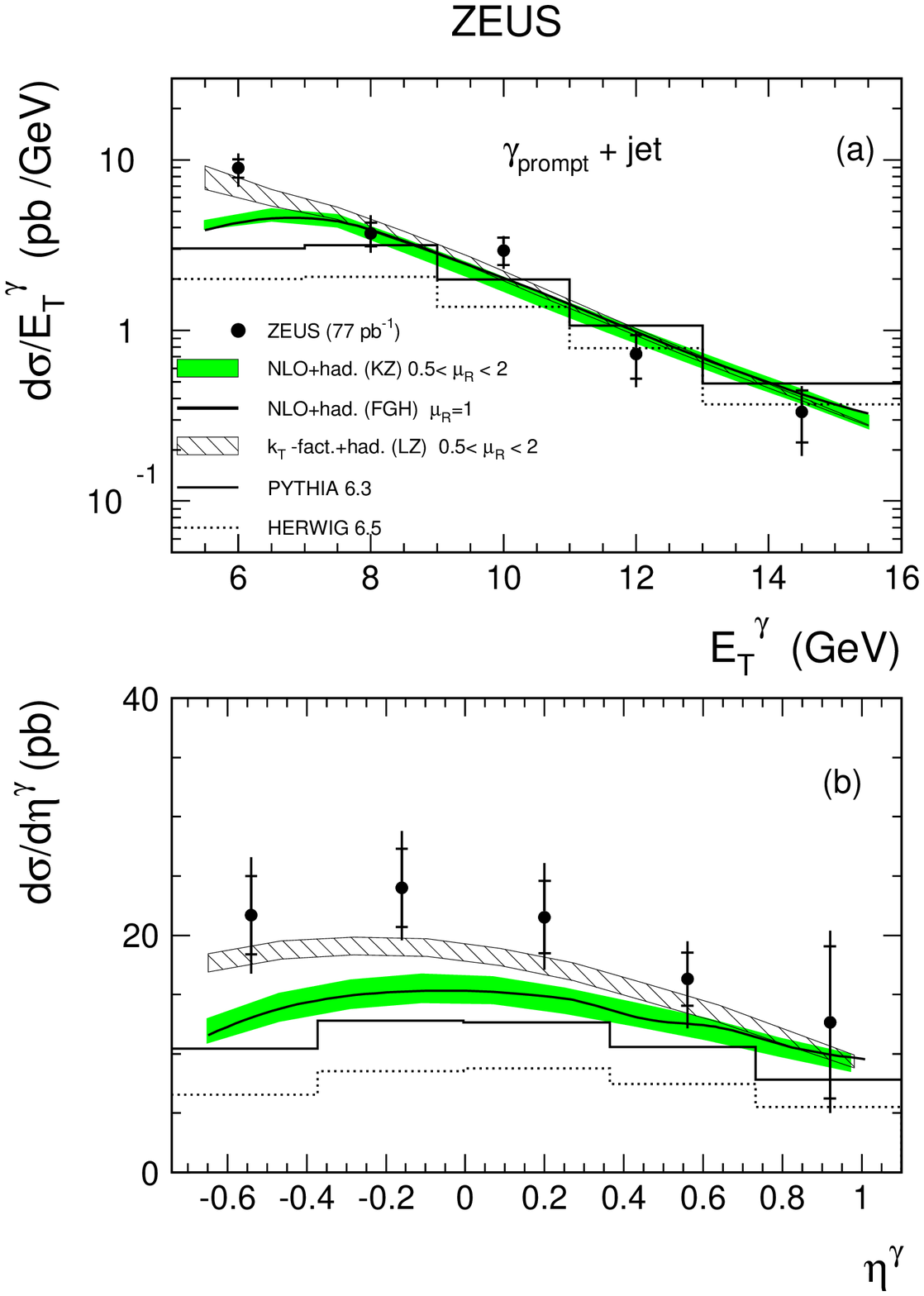}
\end{minipage}\hspace*{2mm}
\begin{minipage}[h]{.49\textwidth}
\includegraphics[height=9.0cm,width=8.0cm,angle=0]{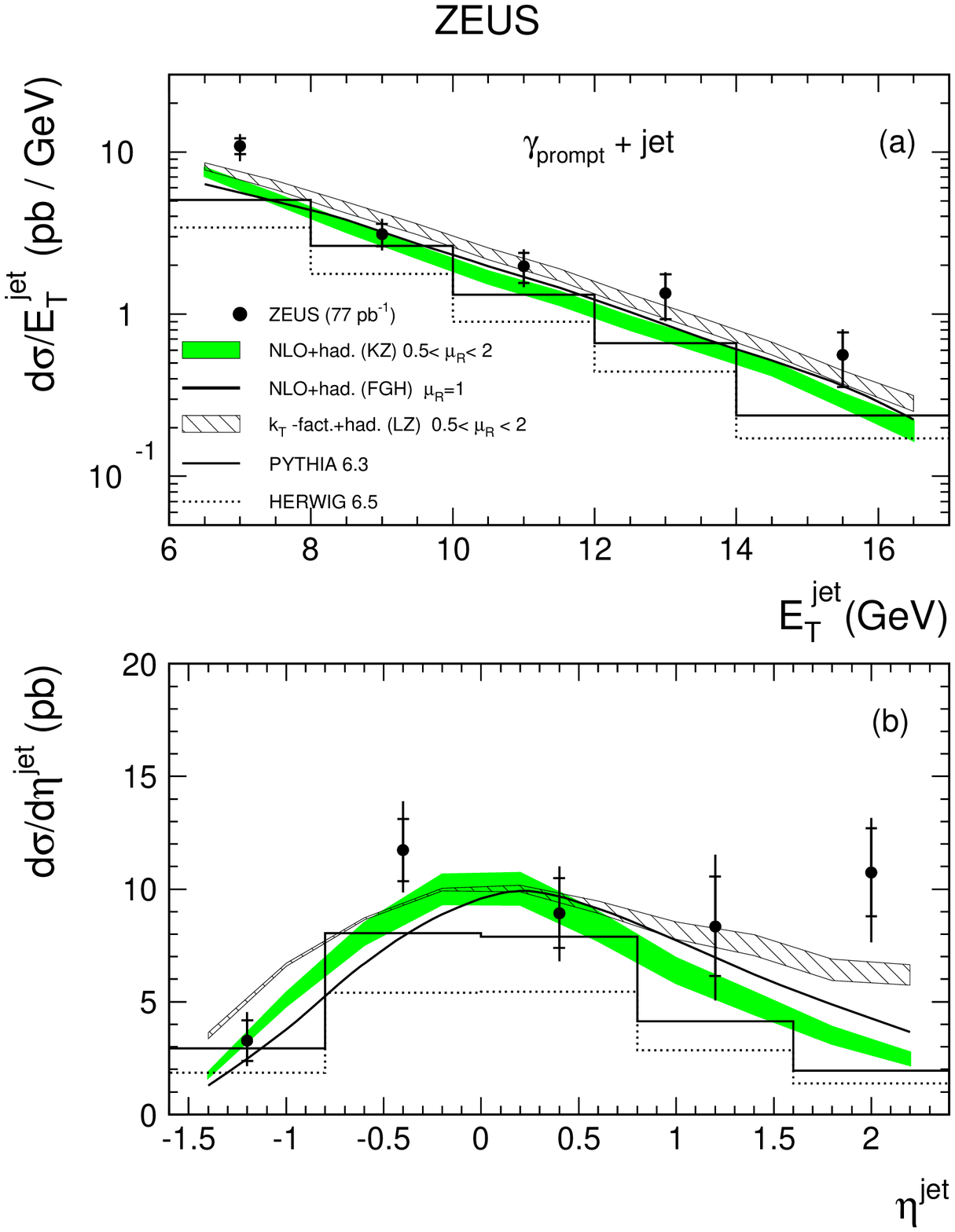}
\end{minipage}
\vspace*{-5mm}
\vspace*{-5mm}
\begin{center}
\caption{\it The differential cross sections for the 
prompt-photon + jet production compared
 to the QCD calculations and MC models [3]}.
\end{center}
\end{figure}
In $ep$ collisions at HERA prompt photons can be produced by one 
of three mechanisms: a direct production, a single resoved 
production and via parton-to-photon fragmentation processes.
The direct process is the Deep Inelastic Compton scattering
on a quark (antiquark): $\gamma q \to \gamma q$.
The single resolved QCD prosesses are $q g \to \gamma q$ and
$q \bar q \to \gamma g$. Photons can be also produced through the
fragmentation of a parton into photon.  However, the  contribution 
of  these fragmentation components is 
significantly reduced (up to 5-6 $\%$) in HERA experiments [2] by 
special isolation criterion. 

In Fig. 1 we show the differential cross sections for the 
prompt-photon events ($E_T^{\gamma} > 5$ GeV) with an 
accompanying 
jet ($ E_T^{\gamma} > 6$ GeV) as functions of
$E_T^{\gamma}$ and $\eta^{\gamma}$ (left panel), $E_T^{jet}$ 
and
$\eta^{jet}$ (right panel) compared to the results of standard QCD 
[14, 15] and $k_T-$factorization calculations (with hadronization
corrections) and MC models (the histograms) taken from Ref. [3].
The shaded bands correspond to the uncertainty in the 
renormalization scale which was changed by a factor of 0.5 and 2.
Fig. 2 shows the distribution for $x_{\gamma}^{obs}$ defined as 
$\Sigma_{\gamma, jet}(E_i - P_Z^i)/(2E_e y)$ (the sum runs over 
the photon candidate and hadron jet).
\begin{figure}[h]
\epsfysize=80mm
\centerline{
\epsfbox{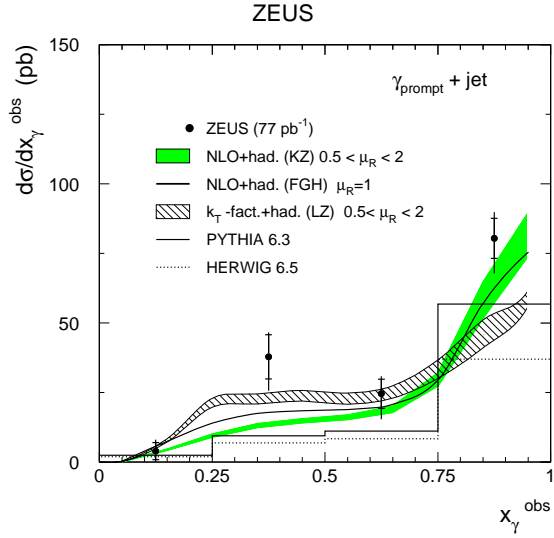}}
\caption{\it The $x_{\gamma}^{obs}$ cross section for the 
prompt-photon production compared to the QCD calculations and 
MC model.}
\end{figure}
We see that the prediction based on the $k_T-$factorization 
approach [12] (corrected for hadronization effects [3]) gives
the best description of the $E_T$ and $\eta$ cross sections.
In particular it describes the lowest  $E_T^{\gamma}$ region
better than the KZ [14] and FGH [15] NLO predictions. The 
$\eta^{jet}$ cross section for the associated jet in the forward   
region and the $x_{\gamma}^{obs}$ distribution in the 
$x_{\gamma}^{obs} < 0.75$ region (the resolved photon 
contribution) are also better reproduced by the our 
calculation. However, it underestimates the observed
cross section at low $E_T^{jet}$, in the forward jet region and
at the $x_{\gamma}^{obs} > 0.75$ (Fig. 2). For the $E_T^{\gamma} 
> 7$ GeV cut (keeping the other cuts the same as before), both
the NLO QCD and the $k_T-$factorization predictions agree
well with the data [3]. The comparison of the the 
$k_T-$factorization predictions and the H1 data [4] was done
in our paper [12].
  
\section{Prompt photon hadroproduction}
Experimenatal data for the inclusive prompt photon 
hadroproduction $p + \bar p \to \gamma + X$ come from
both the D$\oslash$ [5] and CDF [6] collaborations.
\begin{figure}
\hspace*{0mm}
\begin{minipage}[h]{.49\textwidth}
\includegraphics[height=6.0cm,width=10.0cm,angle=0]{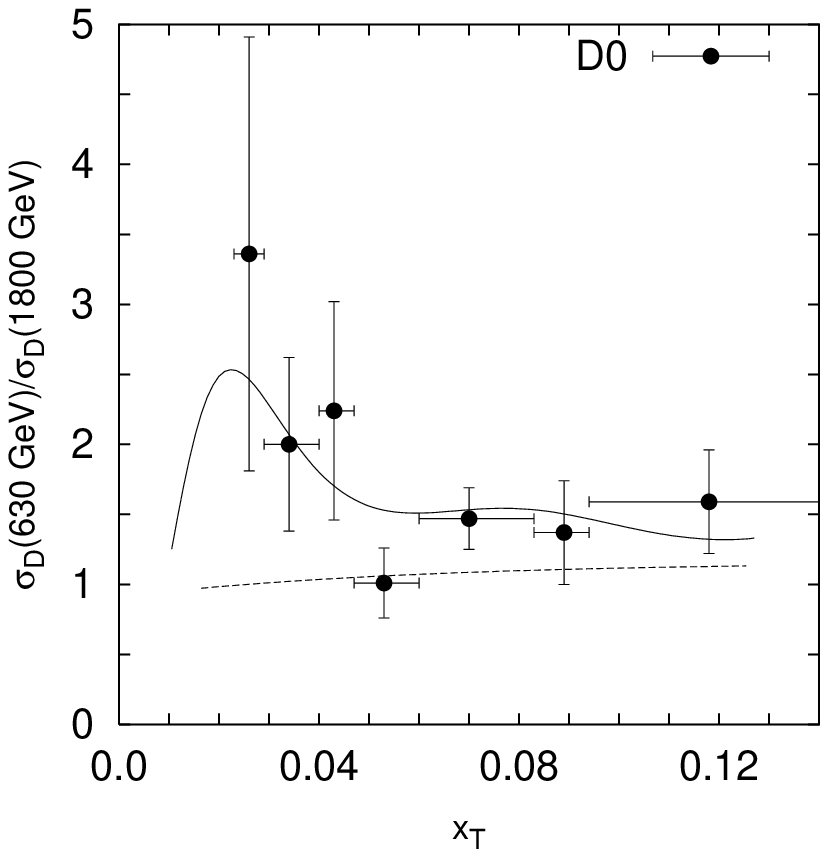}
\end{minipage}\hspace*{2mm}
\begin{minipage}[h]{.59\textwidth}
\includegraphics[height=6.0cm,width=10.0cm,angle=0]{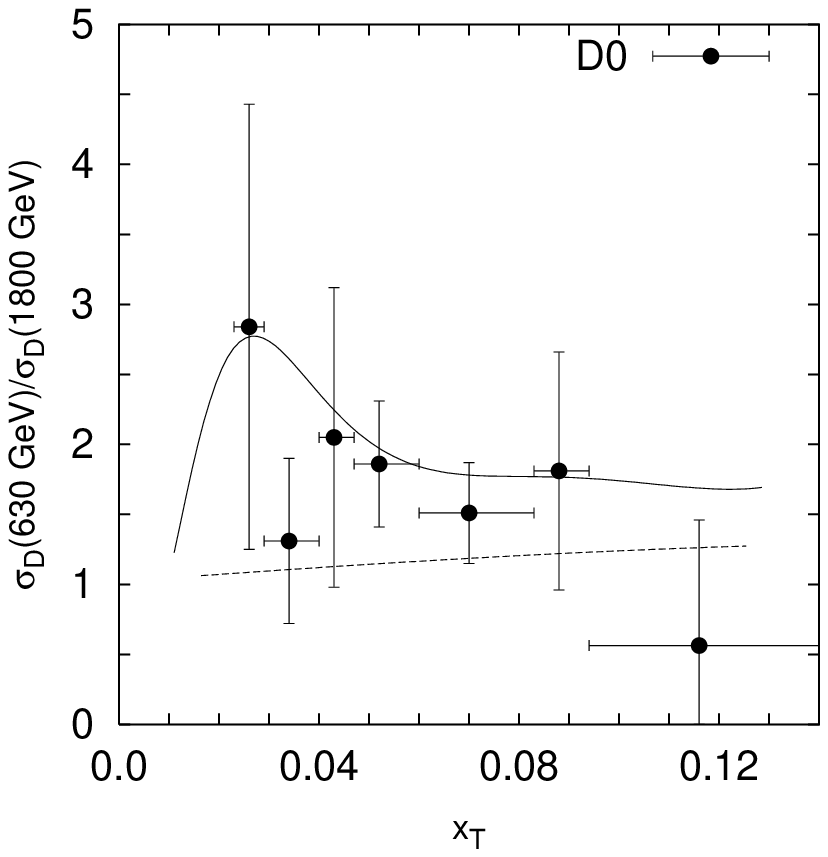}
\end{minipage}
\vspace*{-5mm}
\begin{center}
\caption{\it The ratio $\sigma_D(630$ GeV)/ $\sigma_D(1800$
GeV) as a function of the $x_T$ variable at $|\eta^{\gamma}|
< 0.9$ (left panel) and at $1.6 < |\eta^{\gamma}| < 2.5$ (right
panel). The solid curves are the $k_T-$factorization results,
the dashed curves are the collinear LO QCD results. The data are
 from D$\oslash$ [5].}
\end{center}
\end{figure}
The results of our calculations for the double differential
cross sections $d\sigma/dE_T^{\gamma}d{\eta}^{\gamma}$ 
in comparison with the data were shown in Ref. [13].
We have found that our predictions agree well with
the D$\oslash$ [5] and CDF [6] data both in normalization and 
shape. 
The comparison between the results of NLO calculations and
the CDF experimental data [6] has shown that the NLO ones
agree with the data more qualitatively. So, the shape of the 
measured cross sections is steeper than that of the NLO
predictions [6]. 

Also the disagreement between data and NLO calculations is
visible~[5, 6] in the ratio of the cross sections
at different energies. This quantity is known
as a very informative subject of investigations and
provides more precise test of the QCD calculations. It is because
many factors which affect the absolute normalization partially or 
completely cancel out. In particular, the cross section ratio
provides a direct probe of the matrix elements of the 
hard partonic subprocesses since the theoretical uncertainties 
due to the quark and gluon distributions are reduced.

 The D$\oslash$ collaboration has published the results of 
measurement [5] for the ratio of 630 GeV and 1800 GeV 
dimensionless cross sections
$\sigma_D$ as a function of scaling variable $x_T$. The
measured cross section $\sigma_D$ averaged over azimuth is 
defined as
  $\sigma_D = {(1/2\pi)} (E_T^\gamma)^3 {d \sigma/
 d E_T^\gamma d \eta^\gamma}$.
The ratio $\sigma_D(630\,{\rm GeV})/\sigma_D(1800\,{\rm GeV})$
compared with the D$\oslash$ experimental data~[5]
in different pseudo-rapidity $\eta^\gamma$ regions is shown in 
Fig. 3.
The solid lines represent the $k_T$-factorization
predictions at default scale $\mu = E_T^{\gamma}$. For 
comparison we show also 
the results
of the collinear leading-order (LO) QCD calculations with
the GRV parton densities~[16] of a proton (as a dashed lines).  
Note that when we perform the LO QCD calculations
we take into account the partonic subprocesses $q g \to \gamma 
q$ and $q \bar q \to \gamma g$
and neglect the small fragmentation contributions, as it was 
done in the $k_T$-factorization case. It is clear that although 
the experimental
points have large errors they tend to support the 
$k_T$-factorization predictions. We would like to point out 
again that now sensitivity of our results
to the non-collinear evolution scheme is minimized.
In the collinear approach, the NLO corrections improve the 
description of the data
and then sum of LO and NLO contributions practically coincides 
with our
results at $x_T > 0.05$~[5]. This fact is clear indicates
that the main part of the collinear high-order corrections is
already included at leading-order level in the 
$k_T$-factorization formalism.
Nevertheless, the experimental data at the lowest $x_T$ are
systematically higher~[5] than NLO QCD predictions in both 
central and forward pseudo-rapidity regions, and this
ratio is difficult to reconcile with the NLO QCD calculations~[6].

Now we want to show some predictions for the prompt photon
with associated muon production at Tevatron and for
the differential  cross section $d\sigma/dE_T^{\gamma}$
at LHC [13].

The experimental data for the $\gamma + \mu$ cross section at
Tevatron come from CDF collaboration~[6] taken at 
$|\eta^\gamma| < 0.9$,  $p_T^\mu > 4$ GeV and $|\eta^\mu| < 1.0$.
\begin{figure}
\hspace*{0mm}
\begin{minipage}[h]{.49\textwidth}
\includegraphics[height=6.0cm,width=10.0cm,angle=0]{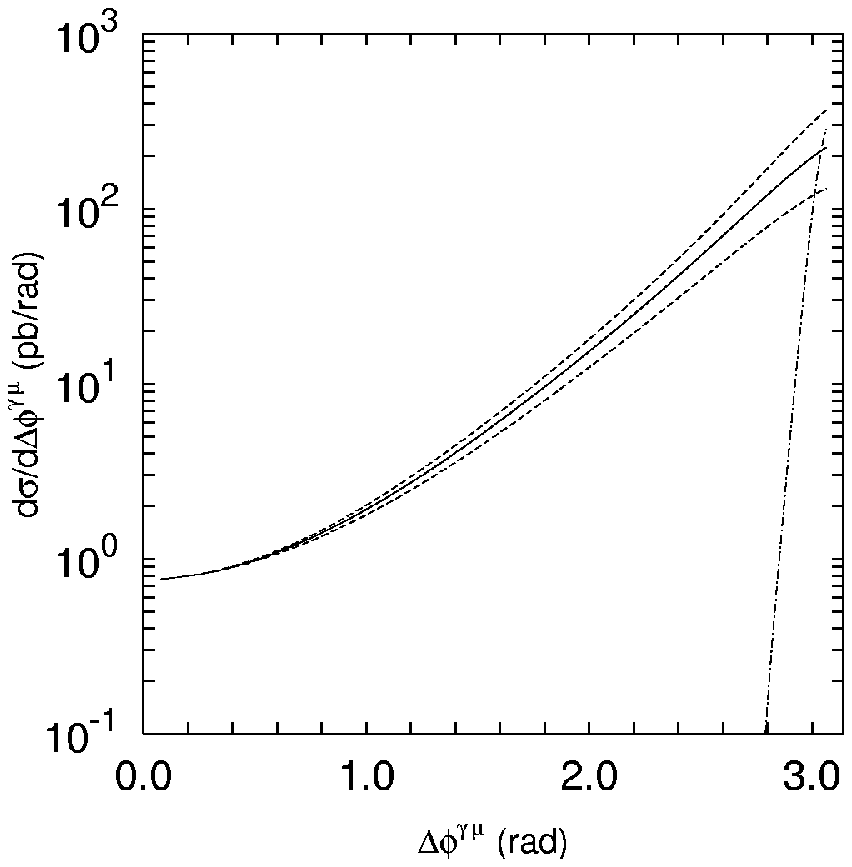}
\end{minipage}\hspace*{2mm}
\begin{minipage}[h]{.59\textwidth}
\includegraphics[height=6.0cm,width=10.0cm,angle=0]
{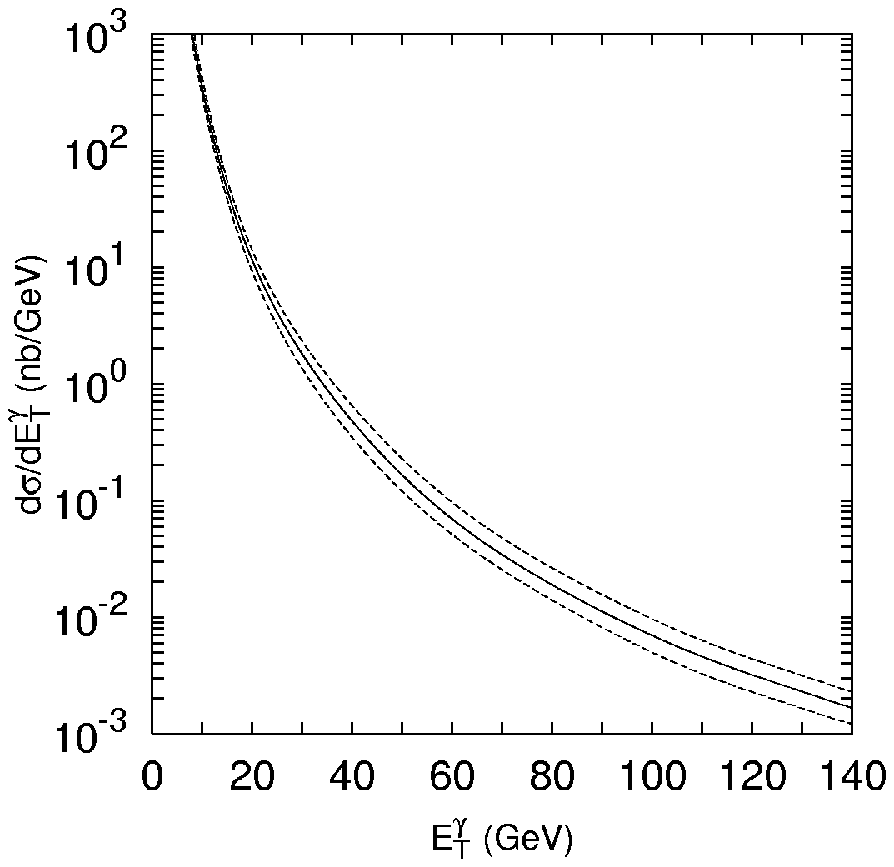}
\end{minipage}
\vspace*{-5mm}
\begin{center}
\caption{\it Left panel: azimutal correlations in associated 
prompt photon 
and muon hadroproduction at Tevatron. Right panel: the 
differential cross section $d\sigma/dE_T^{\gamma}$ for
 inclusive prompt photon hadroproduction at LHC: 
$|\eta^{\gamma}| < 2.5, \sqrt s = 14$ TeV.}
\end{center}
\end{figure}
The transverse momentum distribution $d\sigma/d p_T^\gamma$
in comparison to experimental data~[6] was shown in Ref. [13].
It was shown  the shape of this distribution
is well described by our calculations, but the
theoretical results slightly overestimate the data
in absolute normalization.  However, in general
the experimental points still lie within theoretical scale 
uncertainties (about 30\%) of our calculations.
 It is important also that our
predictions practically coincide with the results of
collinear NLO QCD calculations~[17], which are much larger than 
LO ones~[6].

Further understanding of the process dynamics
and in particular of the high-order effects
may be obtained from the angular correlation between the
transverse momenta of the final state particles~[18]. 

 The differential cross section 
$d\sigma/d\Delta\phi^{\gamma \mu}$
calculated at $p_T^\mu > 4$ GeV, $|\eta^\mu| < 1.0$ and 
$|\eta^\gamma| < 0.9$ is shown in Fig.~4 (left panel).
The solid curve corresponds to the default scale $\mu =
E_T^{\gamma}$, wheres upper and lower dashed curves correspond
to the $\mu = E_T^{\gamma}/2$ and $\mu =2E_T^{\gamma}$
scales. The result of LO QCD calculations
also shown (as a dash-dotted curve). 
One can see a striking difference in shape between
$k_T$-factorization results and collinear LO QCD ones.
The predictions of the NLO QCD calculations for this
distribution are still unknown.
Fig. 4 (right panel) shows also our prediction for the 
differential cross section $d\sigma/dE_T^{\gamma}$ of inclusive
prompt photon production at LHC. 
The direct comparison between NLO calculations, our results
and the experimental data
should give a number of interesting insights.

In summary, we have shown that the $k_T-$factorization 
approach
describes the prompt photon production at HERA
better than standard NLO calculations at certain
cuts for transverse energy of observable photon and jet.
We have found that  the $k_T-$factorization approach
gives specific prediction for the ratio of two cross sections  
calculated at two Tevatron energies. It provides a direct probe of 
the off-shell mass matrix elements. It means that futher
experimental and theoretical investigations promise us a number
exciting insights.
 
 The authors thank S. Chekanov for discussion of the ZEUS data. 
N.Z. thank the Organizing Committee for the support and
hospitality. A.V.L. was support in part by the President grant 
(MK-9820.2006.2). Also this research was support by the FASI of 
Russia (grant NS-8122.2006.2).


\begin{thebibliography}{99}
\bibitem{bib1}
M.M.~Fontannaz, J.Ph.~Guillet and G.~Heinrich, Eur. Phys. J. {\bf 
C21}, 303 (2001).
\bibitem{bib2}
J.~Breitweg {\sl et al.} (ZEUS Collab.), Phys. Lett. {\bf 
B413}, 201 (1997), Phys. Lett. {\bf B472}, 175 (2000);
 S.~Chekanov {\sl et al.} (ZEUS Collab.), 
Phys. Lett. {\bf B511}, 19 (2001).
\bibitem{bib3}
S.~Chekanov {\sl et al.} (ZEUS Collab.), DESY 06-125, 
hep-ex/0608028.
\bibitem{bib4} A.~Aktas {\sl et al.} (H1 Collab.),
Eur. Phys. J. {\bf C38}, 437 (2005).
\bibitem{bib5}
B.~Abbott {\sl et al.} (D$\oslash$ Collab.), Phys. Rev. Lett.
{\bf 84}, 2786 (2000);\\
V.M. Abazov {\sl et al.} (D$\oslash$ Collab.), Phys. Rev. Lett.
{\bf 87}, 251805 (2001).
\bibitem{bib6}
D. Acosta {\sl et al.} (CDF Collab.), Phys. Rev. {\bf D65},
112003 (2002);  Phys. Rev. {\bf D70}, 032001 (2004);
T. Affolder {\sl et al.} (CDF Collab.), Phys. Rev. {\bf D65},
012003 (2002).
\bibitem{bib7}
J. Huston {\sl et al.} (CTEQ Collab.), Phys. Rev. {\bf D51}, 6139 
(1995).
\bibitem{bib8}
L.~Apanasevich {\sl et al.}, Phys. Rev. {\bf D59}, 074007 (1999). 
\bibitem{bib9}
 H.-L.~Lai and H.-N.~Li, Phys. Rev. {\bf D58}, 114020 (1998).
\bibitem{bib10}
A.~Kumar {\sl et al.}, Phys. Rev. {\bf D68}, 014017 (2003).
\bibitem{bib11}
M.A.~Kimber, A.D.~Martin, M.G.~Ryskin, Eur. Phys. J. {\bf C12}, 
655 (2000).
\bibitem{bib12}
A.V. Lipatov, N.P. Zotov, Phys. Rev. {\bf D72}, 054002 (2005).
\bibitem{bib13}
A.V. Lipatov, N.P. Zotov, J. Phys. G.: Nucl. Part. Phys. {\bf 
34}, (2007).
\bibitem{bib14}
M. Krawczyk, A. Zembrzuski, Phys. Rev. {\bf D64}, 14017 (2001);
hep-ph/0309308.
\bibitem{bib15}
M. Fontannaz, J.P. Guillet, G. Heinrich, Eur. Phys. J. {\bf C21},
303 (2001);\\
 M. Fontannaz, G. Heinrich, Eur. Phys. J. {\bf C34}, 191 (2004).
\bibitem{bib16}
M.~Gl\"uck, E.~Reya and A.~Vogt, Phys. Rev. {\bf D46}, 1973 
(1992); Z. Phys. {\bf C67}, 433 (1995).
\bibitem{bib17} B.~Bailey, E.L.~Berger and L.E.~Gordon, Phys. Rev. 
{\bf D54}, 1896 (1996).
\bibitem{bib18}
S.P.~Baranov, N.P.~Zotov, A.V.~Lipatov, Phys. Atom.
Nucl. {\bf 67}, 824 (2004).
 
\end{thebibliography}
\end{document}